\documentclass{bioinfo}
\copyrightyear{2015}
\pubyear{2015}

\begin{document}
\firstpage{1}

\title[BioNetGen 2.2]{BioNetGen~2.2: Advances in Rule-Based Modeling}

\author[Harris \textit{et~al}]{
Leonard A.\ Harris\footnote{Present address: Department of Cancer Biology, Vanderbilt University School of Medicine, Nashville, TN, USA}\,\,, 
Justin S.\ Hogg,
Jos\'{e}-Juan Tapia,
John A.~P.\ Sekar,
Sanjana A.\ Gupta,
Ilya Korsunsky\footnote{Present address: Department of Computer Science, Courant Institute of Mathematical Sciences, New York University, New York, NY and The Feinstein Institute for Medical Research, Manhasset, NY, USA}\,\,, 
Arshi Arora\footnote{Present address: Department of Epidemiology and Biostatistics, Memorial Sloan-Kettering Cancer Center, New York, NY, USA}\,\,,
Dipak Barua\footnote{Present address: Department of Chemical and Biochemical Engineering, Missouri University of Science and Technology, Rolla, MO, USA}\,\,, 
Robert P.\ Sheehan, 
and James R.\ Faeder\footnote{To whom correspondence should be addressed.}
}

\address{Department of Computational and Systems Biology, University of Pittsburgh School of Medicine, Pittsburgh, PA, USA}

\history{Received on XXXXX; revised on XXXXX; accepted on XXXXX}

\editor{Associate Editor: XXXXXXX}

\maketitle

\begin{abstract}

\section{Summary:}
\mbox{BioNetGen} is an open-source software package for rule-based modeling  of complex biochemical systems. Version~2.2 of the software introduces numerous new features for both model specification and simulation. Here, we report on these additions, discussing how they facilitate the construction, simulation, and analysis of larger and more complex models than previously possible.

\section{Availability:}
Stable \mbox{BioNetGen} releases (Linux, Mac~OS/X, and Windows), with documentation, are available at \href{http://bionetgen.org}{http://bionetgen.org}. Source code is available at \href{http://github.com/RuleWorld/bionetgen}{http://github.com/RuleWorld/bionetgen}.

\section{Contact:}
\href{bionetgen.help@gmail.com}{bionetgen.help@gmail.com}

\section{Supplementary information:} 
Supplementary materials are available at \textit{Bioinformatics\/} online.

\end{abstract}

\section{Introduction}\label{sec:intro}

\begin{figure}[!tbp] 
\centering
\includegraphics[width=0.45\textwidth]{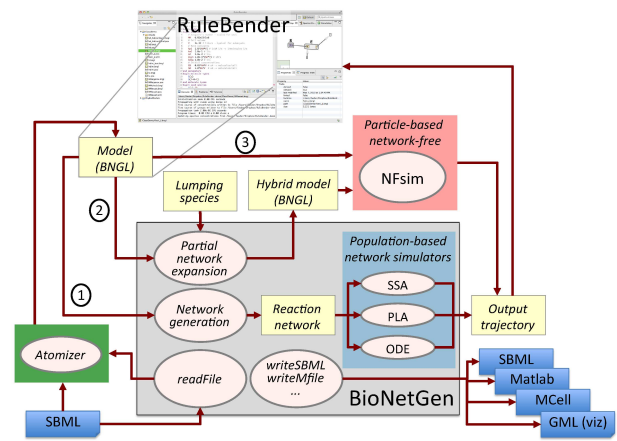}
\caption{Basic overview of the RuleBender/BioNetGen/NFsim software stack. BNGL models may be constructed and edited in RuleBender as well as translated from an SBML model using Atomizer (either with the \mbox{BioNetGen} \texttt{readFile} action or the web application at \href{http://ratomizer.appspot.com/}{ratomizer.appspot.com}). Models can then be simulated in three ways: (1) with a built-in population-based network simulator (ODE, SSA, PLA) after explicit network generation; (2) as a hybrid particle/population (HPP) model in NFsim after partial network expansion (``lumping'' species must be provided \citep{Hogg2014}); (3) with NFsim directly. In all cases, output trajectories are passed back to RuleBender for display. BioNetGen also has a number of built-in methods for exporting to third-party formats, such as SBML and Matlab. (ODE: ordinary differential equations; SSA: stochastic simulation algorithm; PLA: partitioned-leaping algorithm; GML: Graph Modeling Language.)}
\label{fig:RB_BNG_NFsim}
\end{figure}

Rule-based modeling is an approach for addressing combinatorial complexity in models of biochemical systems \citep{Chylek2014, Chylek2015}. Instead of manually enumerating all possible species and reactions that can exist within a system, a rule-based model defines only the reactive motifs within macromolecular complexes and the interactions and modifications that involve those motifs. \mbox{BioNetGen} is an open-source software package for constructing and simulating rule-based models \citep{Blinov2004,Faeder2009}. It has been used to model a variety of biological processes, including cell signaling, gene regulation, and metabolism \citep[and references therein; see also \href{http://bionetgen.org/index.php/Model\_Examples}{bionetgen.org/index.php/Model\_Examples}]{Chylek2014, Chylek2015}. Models are written in a human-readable, text-based modeling language known as BNGL (\mbox{BioNetGen} language). Numerous user-specified actions can be added to a BNGL model file, including generating a reaction network and performing deterministic or stochastic simulations. Models can also be exported to different formats, such as SBML \citep{Hucka2003} and \mbox{MATLAB} language (The MathWorks Inc., Natick, MA). Furthermore, \mbox{BioNetGen} interfaces with \mbox{NFsim} \citep{Sneddon2011}, a ``network-free'' simulator that avoids enumeration of species and reactions, which may be intractable for large models. \mbox{RuleBender} is a visual interface for \mbox{BioNetGen}, with features that include syntax checking\,/\,highlighting and visualizations for model debugging and comparison \citep{Xu2011, Wenskovitch2014}. \mbox{BioNetGen} is also used as a  network generator and simulator in a number of third-party tools including the Virtual Cell \citep{Moraru2008}, BioUML \citep{Kolpakov2006}, SRSim \citep{Grunert2011}, Parts \& Pools \citep{Marchisio2014}, pySB \citep{Lopez2013}, and BioNetFit \citep{Thomas2015}.

Taken together, the \mbox{BioNetGen}\,/\,\mbox{NFsim}\,/\,\mbox{RuleBender} suite of tools (Fig.~\ref{fig:RB_BNG_NFsim}) provides powerful capabilities for rule-based modeling and simulation. Until recently, however, there were several significant shortcomings: (i) BNGL models were limited to mass-action kinetics; (ii) network-based stochastic simulations were limited to computationally expensive ``exact" methods; (iii) network-free simulations were limited to fewer than a few million particles due to high memory load; (iv) there was no facility for importing models from SBML format. Each of these shortcomings has been addressed in the recent \mbox{BioNetGen} 2.2.x series of releases, as described below.
%
%
Additional information can be found in extensive online documentation at  \href{http://bionetgen.org/index.php/Documentation}{http://bionetgen.org/index.php/Documentation} (see Supplementary Information for specific links).

\section{Additions}

\subsection{Functional rate laws}\label{sec:functions}

\mbox{BioNetGen}~2.2 introduces the ability to define arbitrary mathematical functions that can be used as rate laws. Functional rate laws are integrated into the BNGL grammar and are supported by both NFsim and the network-based simulators (deterministic and stochastic) included in the \mbox{BioNetGen} package. Functions are defined using ``observables,'' which compute the concentrations of species with specified properties \citep{Faeder2009}, and can either be evaluated  \emph{globally\/}, i.e., over the entire system, or \emph{locally\/}, over a specified molecule or molecular complex. Local functions greatly expand the expressiveness of BNGL because they enable a single rule to specify many reactions with rates that depend on the specific properties of the reacting species \citep{Sneddon2011}. 

\subsection{Accelerated-stochastic simulation}\label{sec:RKPLA}

$\tau$\/~leaping is an approach for accelerating stochastic simulations of biochemical networks \citep{Gillespie2007}. Numerous variants of $\tau$\/~leaping have been proposed in the literature, including a multiscale variant known as the partitioned-leaping algorithm (PLA) \citep{Harris2006}. In their simplest realizations, $\tau$\/-leaping methods are analagous to the explicit forward Euler method for solving ordinary differential equations (ODEs). Therefore, as with ODE integrators, higher-order and implicit versions of $\tau$\/~leaping algorithms are possible. \mbox{BioNetGen}~2.2 includes an explicit Runge-Kutta implementation of the PLA (RK-PLA) that can be used on any model for which a reaction network can be generated. Additional information, including a performance analysis, is provided in the Supplementary Information.

\subsection{Hybrid particle/population simulation}\label{sec:HPP}

Network-free simulation methods that do not enumerate species and reactions are often required to simulate complex models \citep{Chylek2014, Chylek2015}. However, they are limited by the fact that memory usage increases linearly with the number of particles. \citet{Hogg2014} introduced a hybrid simulation method that treats a user-defined set of species as population variables rather than as particles. This hybrid particle/population (HPP) approach avoids the memory costs associated with having large pools of identical particles and was shown to significantly reduce computational memory expense with no effect on simulation accuracy and little effect on run time \citep{Hogg2014}. HPP is implemented within \mbox{BioNetGen}~2.2 and can be run using NFsim version~1.11 or later.

\subsection{SBML-to-BNGL translation}\label{sec:SBML2BNGL}

SBML is a widely-used model exchange format in systems biology \citep{Hucka2003}. Models encoded in SBML are \emph{flat}, i.e., their species do not have internal structure, which limits their utility for rule-based modeling. \mbox{BioNetGen}~2.2 includes an SBML-to-BNGL translator, 
called Atomizer (also available as a web tool at \href{http://ratomizer.appspot.com/}{ratomizer.appspot.com}), that can extract implicit molecular structure from flat species \citep{Tapia2013}. A full report on Atomizer and its application to the \mbox{BioModels} database \citep{Li2010} is currently in preparation. However, \citet{Tapia2013} reported that an early version of the tool could recover  implicit structure for about 60\% of species in models within the database that contain ${\geq}20$ species. Thus, Atomizer makes available a large set of pre-existing models in a rule-based format, facilitating their visualization \citep{Wenskovitch2014} and extension \citep{Chylek2015}.

\subsection{Additional features}

\mbox{BioNetGen}~2.2 also introduces a number of additional features to those described above, including a null symbol (`\texttt{0}') that can act as a source or a sink in rules and new actions for performing parameter scans, generating MATLAB Executable (MEX) files, exporting to formats readable by third-party simulators (e.g., MCell \citep{Kerr2008}), and outputting graphical visualizations at different scales \citep{Sekar2015}. We have also developed  parameter estimation tool called {\tt ptempest} (available at \href{http://github.com/RuleWorld/ptempest}{http://github.com/RuleWorld/ptempest}) that combines Bayesian Monte Carlo methods\citep{Klinke2014,Eydgahi2013} with parallel tempering \citep{Earl2005} to accelerate sampling. Many new simulation options have also been added, e.g., to allow simulation arguments to be loaded from file and to terminate a simulation if a user-defined logical condition is met. A comprehensive listing of all new actions and arguments added in \mbox{BioNetGen} 2.2 is provided in the Supplementary Information (see also \href{http://bionetgen.org/index.php/BNG\_Actions\_Args}{http://bionetgen.org/index.php/BNG\_Actions\_Args}).

\section{Conclusion}

\mbox{BioNetGen} is an active, open-source project that highly encourages contributions from interested members of the modeling community. Contact information, as well as numerous links to online documentation, is provided in the Supplementary Information. Ongoing development efforts include interfacing with spatial simulators \citep{Sullivan2015}, free-energy-based modeling \citep{Hogg2013Thesis}, and improved support for community standards (e.g., SBML-multi, SED-ML). Recent independent efforts that leverage the \mbox{BioNetGen} framework include a standard for annotation of rule-based models \citep{Misirli2016} specified in BNGL or Kappa \citep{Danos2007a}, another widely-used rule-based modeling language, and \mbox{BioNetFit} \citep{Thomas2015}, a parameter estimation tool for models simulated using either \mbox{BioNetGen} or \mbox{NFsim}.

\section*{Acknowledgements}

We thank 
Robert Clark,
Thierry Emonet,
Robert Engelke,
William Hlavacek,
Cihan Kaya,
G. Elisabeta Marai,
Nikketh Nair,
Daniel Packer,
James Pino,
Adam Smith,
Michael Sneddon,
Lori Stover,
Joseph Vigil
\& John Wenskovitch
for technical help and useful discussions.

\paragraph{Funding\textcolon} 
NIH grants P41 GM103712, R01 AI107825, R01 GM115805, and P01 	HL114453 and NSF Expeditions in Computing Grant (award 0926181). JSH and RPS received support through T32 EB009403.

\bibliographystyle{bioinformatics}
%
%

\end{document}